 \documentclass{caosp309}

\usepackage{graphicx}
\usepackage{natbib}
\usepackage{amsmath}	
\usepackage{amssymb}	

\usepackage{threeparttable}
\usepackage{rotating}
\usepackage{pdflscape}
\usepackage{longtable}
\usepackage[font=normalsize,labelfont=bf,labelsep=space]{caption}
\DeclareUnicodeCharacter{2212}{-}

\articleNo{?}
\pubyear{2022}
\volume{?}
\volnumber{?}
\firstpage{?}
\received{?, 2022}
\accepted{?, 2022}

\begin{document}

\hauthor{D.\,Zengin \c{C}amurdan and B.\,\"{O}zkarde\c{s}}
\title{Detection of a cyclic period change in the contact binary TU UMi}

\author{
        D.\,Zengin \c{C}amurdan\inst{1}\orcid{0000-0003-2596-1775}
      \and
        B.\,\"{O}zkarde\c{s}\inst{2,3}   
       }

\institute{
           Ege University, Science Faculty, Department of Astronomy and Space Sciences, 35100 Bornova,Izmir, Turkey, \email{dicle.zengincamurdan@ege.edu.tr}
         \and 
           Department of Space Science and Technologies, Faculty of Arts and Sciences,
 \c{C}anakkale Onsekiz Mart University, Terzio\u{g}lu
 Kamp\"{u}s\"{u}, TR-17020, \c{C}anakkale, Turkey,
         \and
           \c{C}anakkale Onsekiz Mart University, Astrophysics Research Center and Ulup{\i}nar Observatory, TR-17100, \c{C}anakkale, Turkey 
          }

\date{February 7, 2022}

\maketitle

\begin{abstract}
This study aims at timing the eclipses of the binary star TU UMi. The times of minima are taken from the literature, from our observations in April 2004 and from TESS observations between 2019 and 2022. The orbital period analysis of the system indicates that there is a cyclic oscillation with an amplitude of 0.0081d and a period of 9.03 yr, accompanied by a continuous decrease at a rate of $dP/dt =-1.12 \times 10^{-7}\,\mathrm{d}\, \mathrm{yr}^{-1}$. We study the secular evolution of the orbital period of the system and the possibility of the existence of a third companion or the magnetic activity cycle of the primary component in the system.
\end{abstract}

\section{Introduction}
\label{intr}

Variability of TU UMi (BD+76544,  HIP 73047) was detected during the Hipparcos mission (ESA, 1997) with a period of $0^d.188546$. \cite{Duerbeck97} reported the system as a contact binary of EW, or a pulsating star with a doubled period. \cite{Kazarovets99} and \cite{Rodri00} classified the source as a delta Scuti type variable star. However, \cite{Rolland02} observed the system in Str\"omgren filters and showed that TU UMi is outside of the instability strip of delta Sct-type pulsators using a colour-magnitude diagram. 

TU UMi was observed spectroscopically by \cite{Pych04} who reported that TU UMi is a triple system containing a close binary and the light contribution of the third component around the contact binary is $L_3/(L_1+L_2)=1.25\pm0.15$. \cite{Pych04} also determined the period of TU UMi at twice the Hipparcos original period. \cite{Rucinski05} presented the first radial velocity orbit of the system: $K_1=35\pm15$ $kms^{-1}$, $K_2=220\pm20$ $kms^{-1}$ and the radial velocity of for the third companion is $-4.16\pm0.20$ $kms^{-1}$. They calculated the mass function of $(M_1+M_2)sin^{3}i=0.65\pm0.27 M_{\odot}$ and an extremely low mass ratio of $M_2/M_1 = 0.16\pm0.07$ and classified it as an EW/W-type binary. \cite{Kju2010} observed brightness changes of the system in \textit{BVR} filters. They reported that the light curves of the system are asymmetric with steeper decreasing branches and unequal maxima and suggested that the third bright companion may be responsible for these variations. Furthermore, TU UMi was included in the Asteroseismic Target List \citep{Schofield19} of solar-like oscillators to be observed in 2-minute cadence with TESS and some of the fundamental parameters combining the Gaia DR2 and Hipparcos data were given in this study. Although the minima times of the system were reported by many researchers, the \textit{O-C} period variation was only pointed out by \cite{Kju2010}, who analyzed the data from the years 2003-2009. Based on a 6-year eclipsing cycle of TU UMi, \cite{Kju2010} interpreted that the \textit{O-C} values are increasing linearly in time as the consequence of the low precision of the determined period. 

In this paper, we reconsidered all reported minima times in the literature and used the TESS data which were obtained between 2019 and 2022, and the \textit{UBVR} observations that were obtained in 2004 at the Ege University Observatory, as well. The period change was then investigated in detail, based on all available light minimum times and the results are presented. Here, we report on the detection of a long-term period decrease superimposed on a cyclic change in the orbital period of TU UMi and discuss the plausible cause of these period changes. We performed the solution of the light-travel time effect (LITE) via the presence of an additional companion and magnetic activity cycle for TU UMi with new minima times.

\section{Observations and determinations of the times of minima}

Johnson \textit{UBVR} observations of TU UMi in terms of differential magnitudes with respect to the comparison star HD131358 were carried out at the Ege University Observatory (EUO) with the 0.48-m Cassegrain telescope equipped with a high-speed three-channel photon-counting photometer \citep{Kalytis99}. Fainter comparison and check stars (i.e. Comp. and Chk.) were chosen and are listed in Tab.\,\ref{tbl-1}. The observations were performed during 8 nights in 2004 in \textit{UBVR} bands and typical exposure times were 10 s. We obtained a total of 432 measurements in all bands. All differential magnitudes were corrected for atmospheric extinction and heliocentric corrections were applied to all observing times. The amplitudes of variable light are $0.055^m$, $0.056^m$, $0.062^m$, and $0.052^m$ for \textit{UBVR} bands, respectively. Hence, both eclipse depths are shallow and it is difficult to distinguish primary and secondary minima in our observations. The primary eclipse is deeper than the secondary one by up to $0.005^m$. Since our major goal is to obtain differential photometry for program stars, we did not observe photometric standard stars during our runs. 

Tab.\,\ref{tbl-1} gives an observing log for data observed at the Ege University Observatory (EUO). To study orbital period changes, we collected a total of 14 light minimum times. From the new observations, eight light minimum timings were determined by the least-squares fit of the observed data using the K-W method \citep{Kwee56}. Furthermore, The TESS light curves of TU UMi are used to determine the mid-times of primary and secondary eclipses of the system, where we obtain 904 minima times in total, of which 454 are the primary minima and 450 the secondary times of minimum. The minima times data come from observations sector 14, 15, 20, 21, 22, 26, 40, 41, 47, and 48. Forty-five computed minima times using TESS data are included in Tab.\,\ref{tbl-2}, while the rest of them are available upon request. Those individual light minimum times, together with their errors, and the \textit{O-C} residuals are listed in Tab.\,\ref{tbl-2}.

\section{Analysis of period change}

The eclipse timing data used in the \textit{O-C} analysis of TU UMi covers almost 19 years from March 2003 to February 2022. An \textit{O-C} diagram with all those timings is plotted in Fig. 1 by using the following light element, where $T_{0}=2458696.10801$ (HJD) and $P=0^d.377088$  are the conjunction time obtained in this study from TESS observations and the orbital period taken from \cite{Rucinski05}, respectively. 

The \textit{O-C} diagram including minimum times using the TESS data is shown in Fig. 1. As it can be seen in the figure, a systematic deviation from the linear ephemeris is present in the current \textit{O-C} diagram and it has a downward curving parabolic character indicating a period decrease of the system, which was not discovered by the previous \textit{O-C} diagrams \citep{Kju2010}. However, with the current data, one additional structure in the \textit{O-C} diagram came out which can be seen in the residuals from the parabolic approximation in Fig. 1. This additional structure has a cyclic character superimposed on the general quadratic trend. We have quickly tested the significance of this cyclic character by looking at the difference in the values of the sum of squared residuals of the fits to the \textit{O-C} data with only parabolic and parabolic+cyclic approximations. The sum of squared residuals $\Sigma(O-C)^2$ turns out to be 0.021 $\mathrm{day}^2$ and 0.0026 $\mathrm{day}^2$ for only parabolic and parabolic+cyclic fits, respectively. As seen from these values the sum of squared residuals substantially improved when the cyclic variation has been taken into account.

To search for the period of the cyclic variations, a periodogram analysis with PERIOD04 \citep{Lenz05} was carried out for the quasi-sinusoidal variation seen in the residuals of \textit{O-C} after the subtraction of a quadratic term, whose result is displayed in Figure 2. We found a significant peak in the power spectrum mainly located around the frequency of $f_1 = 3.00744 \times 10^{-4}\, \mathrm{d}^{-1}$, which corresponds to a period of 3325 days ($\sim 9.1$ yr). This value is then used as an initial $P_{mod}$ parameter for the non-linear least-squares fitting the \textit{O-C} values.

The cyclic variations are usually explained either by the light travel time effect (LTTE) via the presence of a third body, or by magnetic activity cycles in one or both components because they are late-type stars. To search for the LTTE effect, the calculated the \textit{O-C} values based on all times of minima were fitted by the equation:

\begin{equation}
\begin{split}
 O-C=&O-[T_0+P_{orb} \times E+\frac{1}{2}\frac{dP}{dE} \times E^2 + \\ 
 &\frac{A}{\sqrt{1-e^2\,cos^2\,\omega}} \times \Bigg\{(1-e^2)\frac{sin\,(\nu+\omega)}{(1+e\,cos\,\nu)}   \Bigg\}
 \end{split}
\end{equation}

where
\begin{equation*}
A=\frac{a_{12}\,sin\,i\sqrt{1-e^2\,cos^2\,\omega}}{2.590 \times 10^{10}}
\end{equation*}

is the semi-amplitude of the light-time effect in days. \textit{E}, $P_{orb}$, $T_0$ and $dP/dE$ represent the cycle number, orbital period, the reference epoch for the primary minimum and the rate of the secular period change of the binary star, respectively. $a_{12}$, $i$, $e$ and $\omega$, $2.590 \times 10^{10}$ are the semi-major axis, inclination, eccentricity, longitude of the periastron of the absolute orbit of the center of mass of the eclipsing binary around the three-body system and the speed of light in km/day, respectively. $\nu$ is the true anomaly of the position of the eclipsing binary’s mass center on this orbit and includes $T_{p}$ that it is the epoch of the passage at the periastron of the eclipsing binary’s mass center along its orbit and $P_{mod}$, which is the modulation period of the third body orbit. 

 The \textit{O-C} method was applied in the analysis using a matlab code given by \cite{Zasche09} to determine the eight free parameters (namely $T_0$, $P_{orb}$, $dP/dE$, $P_{mod}$, $T_{p}$ , $a_{12}\,sini$, $e$, $\omega$) by least-squares fitting the \textit{O-C} values taking into account both variations (parabolic+cyclic). The best theoretical fit made to the $O-C$ diagram and the residuals obtained from the fit are shown in Fig. 1 and Fig.3 and the derived parameters and their errors are given in Tab.\,\ref{tbl-3}. The sum of square residuals from the best fit is 0.0026 $\mathrm{days^{2}}$.

\section{Results and conclusions}
\subsection{Mass transfer between the components}

TU UMi is a short-period ($0^\mathrm{d}.3770$) W-type contact binary where the primary component is estimated to be F-type star by \citet{Rucinski05} and is poorly investigated in detail. By analyzing a total of 926 times of light minimum, the orbital period of TU UMi is changing with a cyclic character superimposed on a decreasing structure. The \textit{O-C} curve of the system shows a clear long-term period decrease at a rate of $dP/dt = -1.12 \times 10^{-7}\,\mathrm{d}\, \mathrm{yr}^{-1}$ ($-0.18\, \mathrm{sec}\, \mathrm{century}^{-1}$). The type of variations, i.e., a long-term decrease combined with a cyclic change, is commonly found in W UMa-type stars that are shown in Tab.\,\ref{tbl-4} for some of the W-type contact binaries between $0.2\, \mathrm{d} < P_{orb} < 0.7\, \mathrm{d}$. Furthermore, \citet{Li2018} studied the light travel-time effect in short-period eclipsing binaries and concluded that the frequency of third bodies found in contact binaries with $P<0.3$ day reaches a value of 0.65 in their samples including 542 eclipsing binaries. Therefore, there may exist a periodic variation superimposed on a secular decrease in the period in the \textit{(O-C)} curve of TU UMi. The long-term period decrease can be explained by mass transfer from the more-massive component to the less-massive one in the case of a conservative mass transfer, or by the angular momentum loss from the system, or by a combination of both mechanisms.

The period investigation of \citet{Kju2010} showed that the \textit{O-C} variation of TU UMi was linear. However, only times of light minima observed between 2003 and 2009 were used in their work. In our \textit{O-C}  diagram, times of light minima obtained by the high accuracy TESS data are used and these data show a downward tendency. Therefore, the TESS data are very important when the period variation of TU UMi is investigated since they will give a quite different result, as shown in Fig. 1 and Fig. 3. Also, a cyclic period change is revealed for the first time in this study. Assuming that the long-term period change of the system can be explained by mass transfer, then some parameters, including the masses of the two components ($M_1=1.313\, M_{\odot}, M_2=0.21\, M_{\odot}$)\footnote{We should state that the absolute values of masses of the system used in this section are taken from \citet{Xu2020} calculated by using equations given in \citet{Gazeas2009} (equations 7–10) where the mass ratio ($q=m_2/m_1$) is about 0.16 (from spectroscopic observation given by \citet{Rucinski05}). }, period, the rate of change of the period of TU UMi were used in the following equation to calculate the rate of mass transfer:

\begin{equation}
\dot{M}=\frac{M_1.M_2}{3(M_1-M_2)}\frac{\dot{P}}{P} ;
\end{equation}

thus, $dM/dt = −0.25\times 10^{-7}M_{\odot} yr^{-1} $ was determined, assuming that the more massive star transfers its present mass to the less massive component on a thermal timescale  $\tau_{th}$ \citep{Paczynski1971} , $\tau_{th}=1.2 \times 10^7 $yr and mass is transferred to the companion at a rate of $M_1/\tau_{th}=1.1 \times 10^{-7} M_{\odot} yr^{-1} $. This value is compatible with the calculated one using the observed period change. This means that mass transfer can describe the parabolic variation.

\subsection{Cyclic variation}
The \textit{O-C} residuals obtained after the subtraction of the quadratic term are shown in Figure 3, whose shape appears to have a quasi-sinusoidal variation with a period of $\sim 9$ yr. This can be explained in two ways: the Applegate mechanism \citep{Applegate92} and the light-time effect. 

\subsubsection{Presence of a possible third component}

One of the possible causes of the cyclic \textit{O-C} variation may be the light travel time effect (LTTE) due to a third body that is physically bound to the system. In order to examine this situation, the LTTE equation given by \cite{Irwin52} (see Equ. 1) was fitted to the \textit{O-C} diagram of the system.

According to the parameters given in Tab.\,\ref{tbl-3}, TU UMi orbits around the triple system’s center of mass in a very low eccentric orbit ($e= 0.05\pm0.01$) with a period of $P_{mod}= 9.03\pm0.02$ years. The projected distance of the center of mass of TU UMi to the center of mass of the three-body system was estimated to be $a_{12}\,sin\,i=1.398\pm0.009$ AU. Using the $P_{mod}$ and $a_{12}\,sin\,i$ values, the mass function of the third body was found to be $f(M_3)=0.03352\pm 0.000001\, M_{\odot} $. The minimum mass of the tertiary ($i = 90^{\circ}$, where $i$ is the inclination of the third body’s orbit) is estimated as $M_3 = 0.611\, M_{\odot}$. According to \citet{Budding2007}'s empirical main sequence table, the spectral type of this companion is estimated to be $\sim$ K3. Therefore, the third body deduced from LTTE may be a low-mass star. In this case, the mass of the third body is comparable to the mass of the secondary component in the binary system. \citet{Pych04} reported that TU UMi is a triple system containing a close binary and about 50 percent of total light coming from the bright tertiary companion in the system. However, this system is also known as a visual binary\footnote{The visual component is WDS 14557+7618.} with a separation of 0.2 arcsec. \citet{Rucinski05} suggest that the third, spectroscopic component is probably identical to the visual component. Apart from this, the third body was not confirmed photometrically and further observational evidence for the existence of a third body needs to be investigated. If the third body is really present in the system, it may have an important role in the evolution of the binary system by drawing angular momentum from the central binary via a Kozai cycle \citep{Kozai62} or a combination of a Kozai cycle and tidal friction.

\subsubsection{Magnetic activity}

The close binary systems with late-type components are well known to be magnetically active, e.g. with chromospheric emission, starspots. The Applegate mechanism suggests that magnetic activity causes a variation of the angular momentum distribution, and then leads to variations in the active component in the system.  So any change in the rotational system of a binary star component due to the magnetic activity will be reflected to the orbit as a consequence of the spin-orbit coupling and the orbital period changes slightly periodically. 

The observed cyclic variation in the \textit{O-C} diagram of TU UMi may be explained as resulting from magnetic activity variations due to star spots. We have calculated the activity related parameters by following the \citet{Applegate92} formulation and assuming the primary component (should be an F2 type dwarf according to the (\textit{B-V}) color index value given by \citet{Duerbeck97}) of the system is responsible from the activity. The parameters are the cycle modulation, period $P_{cyc}=9.03$ years, the amplitude of the cyclic period variation, $\Delta P=1.005$ sec $\mathrm{cycle}^{-1}$, the angular momentum transfer $\Delta J= −7.66\times 10^{47}  \, \mathrm{g}  \, \mathrm{cm}^2  \, \mathrm{s}^{-1}$ required to produce the observed cyclic effect on the orbital period, required energy $\Delta E= 2.28 \times 10^{41}  \, \mathrm{ergs}$ for the $\Delta J$ transfer, the corresponding luminosity change $\Delta L=0.633 \, L_{\odot} $, and, finally the subsurface magnetic field $B = 22.9\,  \mathrm{kG}$ of the primary component. The period variation $\Delta P/P$ can be used for calculating the variation of the quadruple moment $\Delta Q$ and this quantity results in $ 2.15 \times 10^{50}\, \mathrm{g}\, \mathrm{cm}^2$, which is within the limits for active binaries (a range of values from $10^{50}$ to $10^{51}\, \mathrm{g}\, \mathrm{cm}^2$). Since the parameters of the two components of the binary were adopted by \citet{Xu2020}, we can calculate the changes of the luminosity to be $\Delta L_1 = 0.633\, L_{\odot}< L_1\,(3.2\,L_{\odot})$. Obviously, the active primary component can provide enough energy to generate such changes. Therefore, the Applegate mechanism can explain the cyclic period change in TU UMi as well. In addition, we should point out that no unequal difference between the two light maxima (Max I-Max II), or the O’Connell effect was not remarkable in the TESS and our observations. In addition to this, according to \citet{Tran13}, strong spot activity causes an effect on the variations of the minima times. The trends of primary and secondary minima time variations are expected to show an erratic, and anticorrelated to each other behaviour and these have been observed in many W UMa type systems. To search for such a variation in our case, we decided to plot primary and secondary \textit{O-C} times determined from TESS data. As TU UMi was observed by TESS intermittently during the period March 2003 to February 2022, we focus on two consecutive data sets (Sector 21, 22 and Sector 47, 48). As it can be seen in Fig. 4, there are no short-term, anticorrelated \textit{O-C} variations in the primary and secondary curves of TU UMi. The results may suggest that TU UMi has a very weak activity and it may be in an inactive state for decades with no significant spot activities. The Applegate mechanism also supports the long-term light variation and the \textit{O-C} curve formed by the times of minima should have the same cycle length. Unfortunately, we do not have precise enough long-term photometric observations for TU UMi to check such brightness variations. Long-term photometric monitoring can clarify TU UMi's plausible magnetic activity cycle characteristics.

\subsubsection{ Long-term period decrease in W-type W UMa stars}

We calculated the angular momentum loss (AML) via magnetic stellar wind which can be determined by the following equation given by \citet{Brad94} for the long-term decrease:

\begin{equation}
\begin{split}
\dot{P} & \approx 1.1 \times 10^{-8}\,q^{-1}\,(1+q^{2}\,(M_1+M_2)^{-5/3} \\
         & \times k^2 \,(M_1\, R_{1}^4+M_2\, R_{2}^4)\,P^{-7/3}
\end{split}
\end{equation}

where $k^2$ is the gyration constant between 0.07 to 0.15 for solar-type stars. By adopting the value of $k^2 = 0.1$, the rate of the orbital period decrease due to AML can be computed as $dP/dt\sim 2.4 \times 10^{-7} \, \mathrm{d}\, \mathrm{yr}^{-1}$ and the timescale of the period decrease is $(dP/dt)\sim 1.56 \times 10^6 \, \mathrm{yr}$ or $\sim 1.6 $ Myr) which is close to the timescale from observed the period decrease derived from $O-C$ variation ($P/(dP/dt)=2.6$ Myr). This may lead to the conclusion that the long-term period decrease of TU UMi can also be constrained by AML, but as it can be seen in Sec. 4.1, the period decrease from conservative mass transfer cannot be neglected. Therefore, a plausible explanation for the long-term period decrease in TU UMi is AML, or the combination of the two mechanisms (AML and mass transfer). Nevertheless, observations in the next decades are necessary to clarify the true shape of the \textit{O-C} diagram.

\subsubsection{ Long-term period decrease in W-type W UMa stars}

Long-term period decreasing and increasing are common for W UMa-type binaries. This variation is usually due to mass transfer or angular momentum loss, which results in orbital period changes. In this study, the analysis of the orbital period of TU UMi was performed and a long-term period decrease was discovered. In general, it is interpreted as caused by conservative mass transfer from the more massive component to the less massive component. In order to find a relation between the observed period change rate using $O-C$ variation and the mass ratios, we compiled the physical parameters and the observed period change rate ($dP/dt$) for some  W-type contact binaries showing a secular period change plus cyclic variations in the literature in a period range of $0.2 < P_{orb} < 0.7$ d. They are listed in Tab.\,\ref{tbl-4} in the order of the increasing orbital period. The plot of the period change versus the mass ratio rate can be seen in Fig. 4, where the positions of 32 W-type W UMa stars are shown. Based on these parameters, there may exist a correlation (solid line) between the period change, $dP/dt$, and the mass ratio, $q$. We fitted the data using the least-squares method and the fitting result yields the following equation: 

\begin{equation}
dP/dt = -5.82(\pm1.55) \, q +3.60(\pm0.60),  
\end{equation}

where $dP/dt$ is in units of $\times 10^{-7}$ $\mathrm{d}\, \mathrm{yr}^{-1}$. This relation tells us that the secular periods of systems are decreasing with the increasing mass ratio. It seems that the mass transfer between the two components may contribute to changes in the mass ratio of a system, which needs to be confirmed using a large sample of W UMa stars. However, in a statistical study for contact binaries by \citet{Latkovic2021}, there has been recently investigated the relation between the orbital period change rate (taking into account both period increasing and decreasing) and the mass ratio for A and W-type systems using a large sample of W UMa stars. They pointed out that there is no correlation between the type of period variability and the mass ratio (see their Fig. 7). In our sample, it seems that the derived relation is more obvious for $q>0.25$. The systems with $q<0.25$ tend to show $dP/dt$ in a large range of $(2-5.3) \times 10^{-7} \, \mathrm{d}\, \mathrm{yr}^{-1}$. The maximum $dP/dt$ value is $3 \times 10^{-7} \, \mathrm{d}\, \mathrm{yr}^{-1}$ for the systems with $q>0.25$. The $q$ value of TU UMi shows that it is in agreement with the observed period change rates of W-type W UMa binaries. Tab.\,\ref{tbl-4} also lists the estimated mass transfer rates of samples, we found them to be in the range $(0.002 - 2)\times 10^{-7}M_{\odot} yr^{-1} $, while TU UMi appears to have a comparable rate of $0.3\times 10^{-7}M_{\odot} yr^{-1}$.  The primary masses of samples ($M_1$) in Tab.\,\ref{tbl-4} apparently show no relationship with the period-change rate ($dP/dt$). However, there is the same relation between the secondary masses ($M_2$) and the period-change rate ($dP/dt$), where $dP/dt$ decreases with increasing $q$. This is understandable because, in the study of \citet{Latkovic2021}, the empirical relationship between the secondary masses and the mass ratio shows a linear correlation, $M_2$ increases with increasing $q$. Most of the systems in Tab.\,\ref{tbl-4} have a confirmed cyclic change related to the magnetic activity of components, or LTTE in the system. Although we accept that the long-term period decrease is due to conservative mass transfer between components, the contribution of angular momentum loss (AML) via magnetic stellar wind, or by a combination of both processes, on the secular period reducing cannot be ruled out. The cyclic \textit{O–C} variation can be explained in terms of the LTTE or solar-like magnetic activity cycle of one of the two components in the system. However, we can not discriminate which mechanism is responsible for the cyclic variation in the \textit{O–C} diagram and it is clear that we need more observations to ascertain that.


\acknowledgements
The authors acknowledge generous allotments of observing time at the Ege University Observatory. We thank S. Evren and G. Ta\c{s} for their support in performing observations at the Ege University Observatory and the staff of the Ege University Observatory. We would like to thank Burak Ula\c{s} for allowing us to use his python code to determine times of minima from TESS observations. And we thank the referee very much for the comments and suggestions that have helped us to improve the manuscript. This paper includes data collected by the TESS mission, which are publicly available from the Mikulski Archive for Space Telescopes (MAST). This research has made use of the SIMBAD online database, operated at CDS, Strasbourg, France, and NASA’s Astrophysics Data System (ADS).

\begin{figure*}
	\includegraphics[width=0.95\columnwidth]{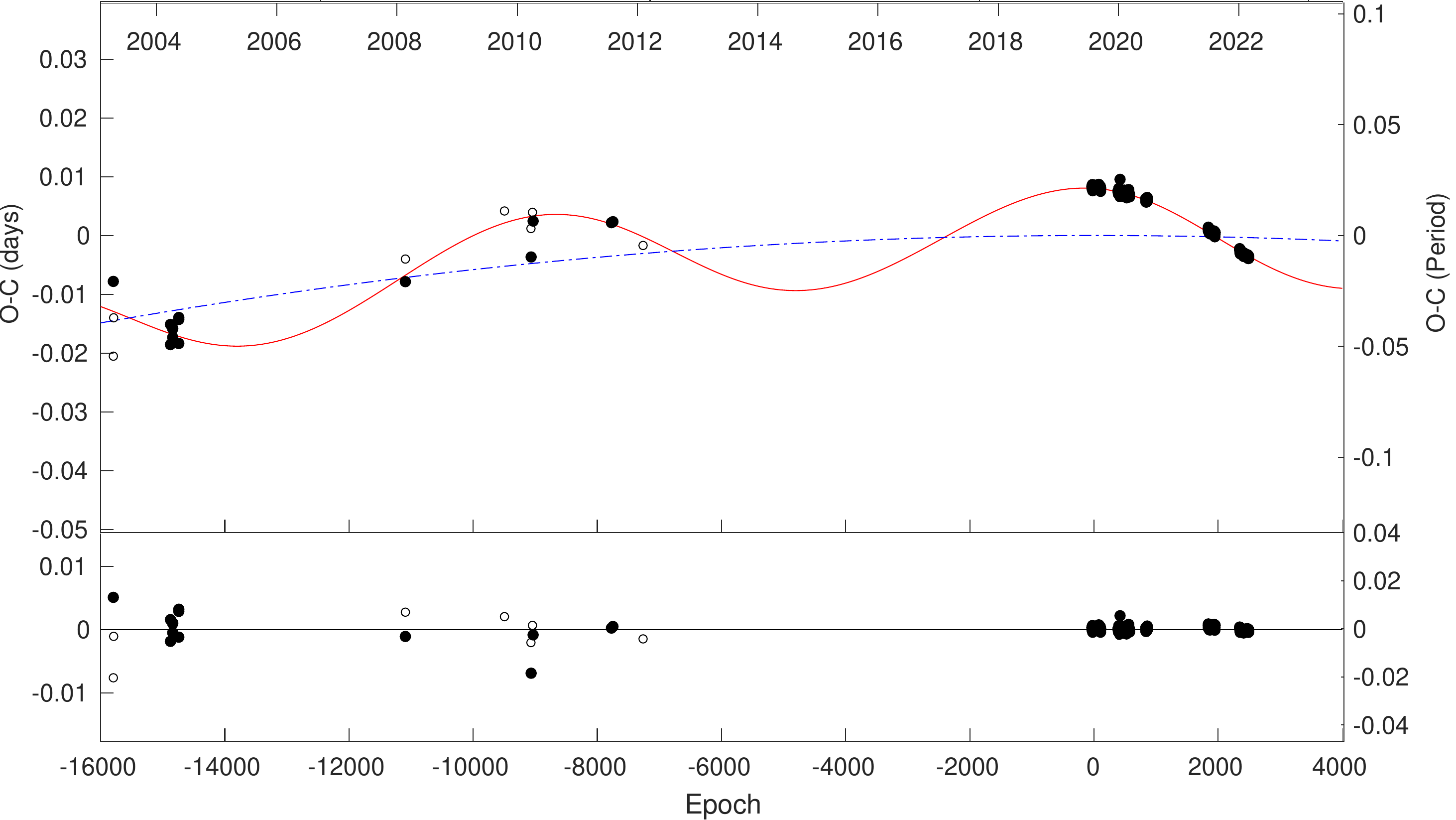}
    \caption{The \textit{O-C} diagram of TU UMi (upper part), where the solid line represents the theoretical LITE variation caused by a 3rd body and the dashed line represents the quadratic ephemeris, and the \textit{O-C} residuals obtained after the subtraction of LITE (lower part).}
    \label{Fig.1 }
\end{figure*}


\begin{table} [h]
\begin{center}
\footnotesize
\caption{Coordinates of the variable, comparison and check stars.\label{tbl-1}}
\begin{tabular}{@{}lcrcl@{}}
\hline
      \hline
Star & $\alpha_{J2000.0}$ & $\delta_{J2000.0}$ & mag (V)\\
\hline
 TU UMi            & 14 55 43.80 & +76 18 23.65 & $8.75\pm0.01$ \\
 HD 131358 (Comp.) & 14 47 10.16 & +76 02 32.13 & $7.38\pm0.01$ \\
 HD 135118 (Chk.)  & 15 06 53.63 & +75 59 07.04 & $8.29\pm0.01$ \\
\hline
      \hline
\end{tabular}

  \end{center}
\end{table}

\begin{figure}[pth!]
	\includegraphics[width=1.0\columnwidth]{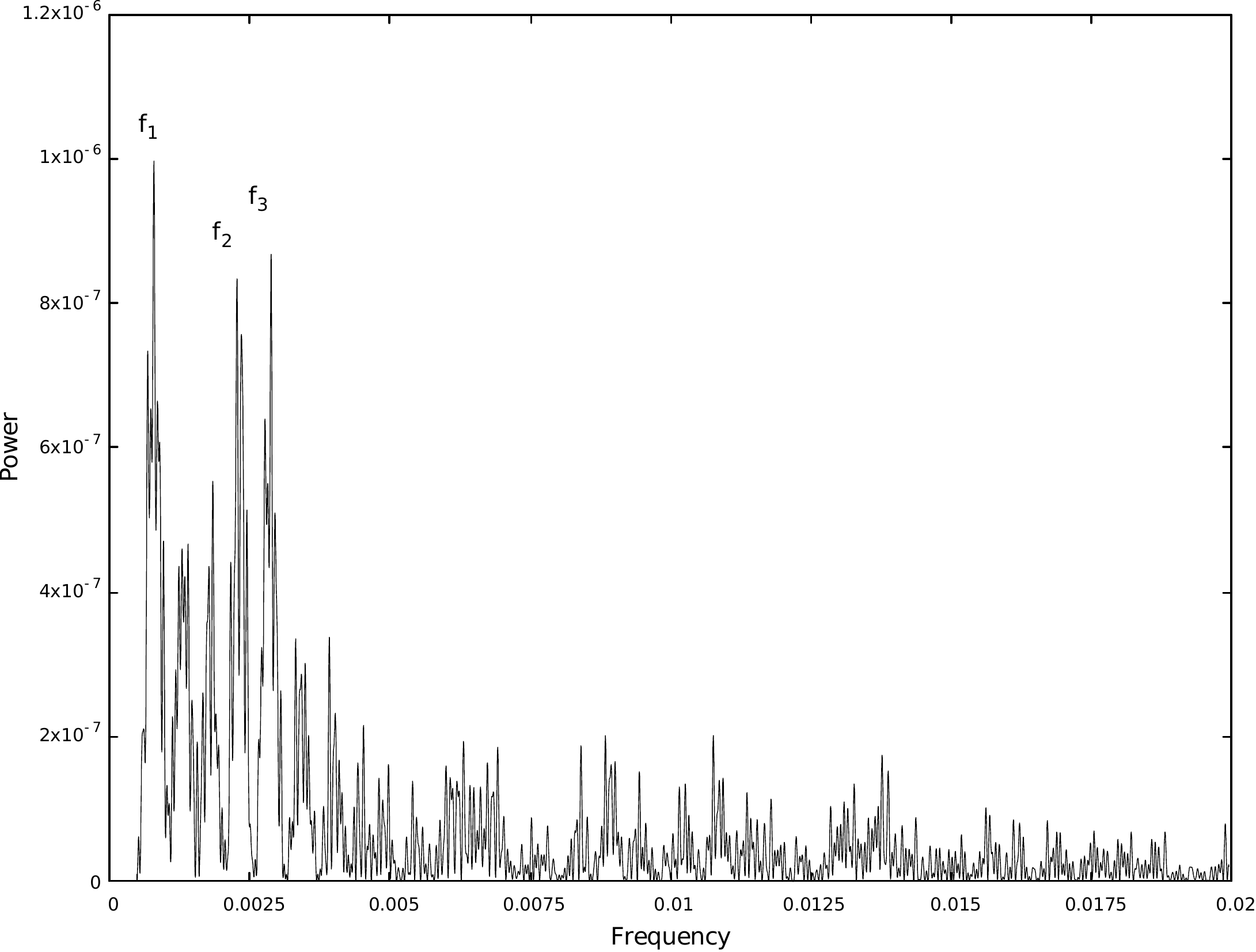}
    \caption{The power spectrum of the residuals from the quadratic fit.}
    \label{Fig.2 }
\end{figure}

\begin{figure}[pbh!]
	\includegraphics[width=1.0\columnwidth]{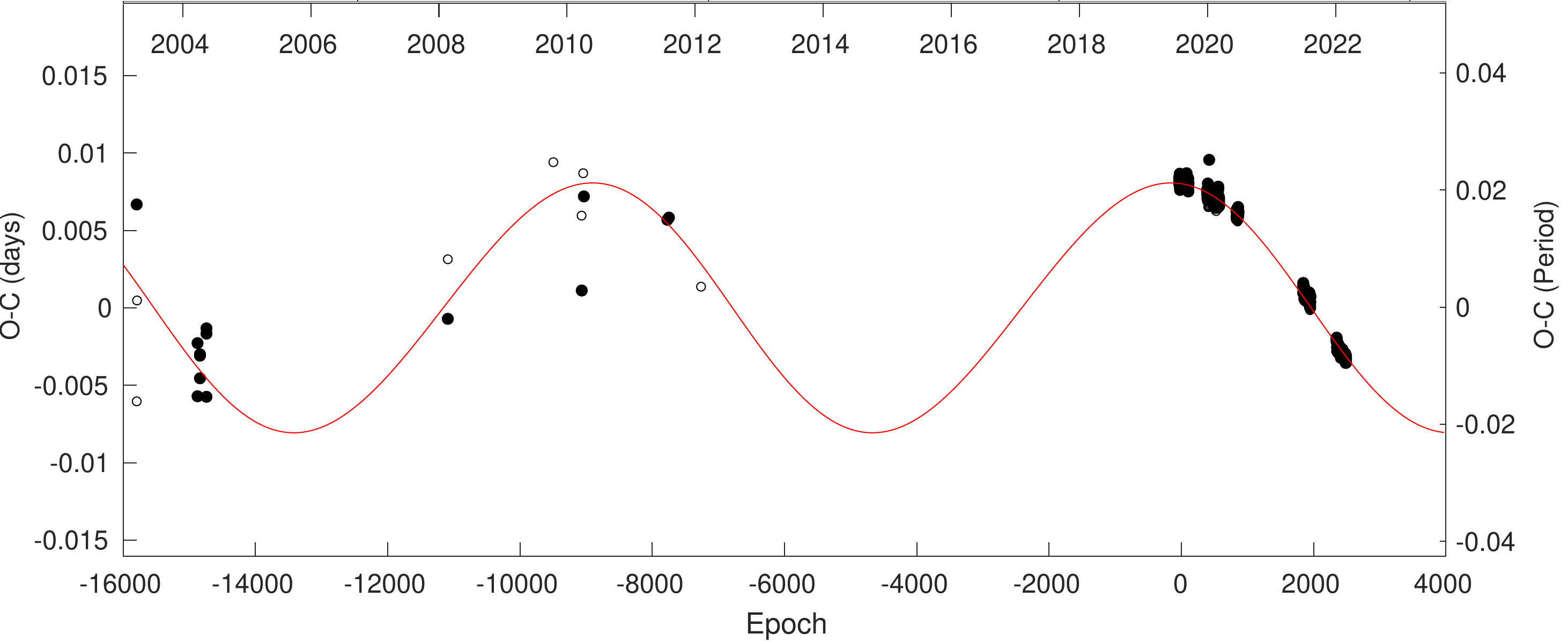}
    \caption{The \textit{O-C} curve of TU UMi shows the residuals from the quadratic fit. The solid line refers to a theoretical orbit of the additional component in the system.}
    \label{Fig.3 }
\end{figure}



\begin{table} 
\begin{threeparttable}
\footnotesize
\caption{Observed times and $\textit{(O-C)}^\mathrm{d}$ values of minimum light for TU UMi.\label{tbl-2}}
\begin{tabular}{ccrcrrrc}
\hline
      \hline
BJD\footnote[1] (2400000+) & Error & Epoch&\textit{$(O-C)_1$}&\textit{$(O-C)_2$}&Residuals&Reference\\
\hline
52739.5938	&	0.0004	&	-15796.0	&	-0.0071	&	0.0067	&	0.0051	&	[1]\\
52739.7696	&	0.0008	&	-15795.5	&	-0.0198	&	-0.0061	&	-0.0076	&	[1]\\
52741.6616	&	0.0003	&	-15790.5	&	-0.0133	&	0.0005	&	-0.0011	&	[1]\\
53086.5050	&	0.0020	&	-14876.0	&	-0.0179	&	-0.0057	&	-0.0019	&	[2]\\
53086.5084	&	0.0022	&	-14876.0	&	-0.0145	&	-0.0023	&	0.0016	&	[2]\\
53100.4599	&	0.0043	&	-14839.0	&	-0.0153	&	-0.0031	&	0.0010	&	[2]\\
53100.4577	&	0.0023	&	-14839.0	&	-0.0167	&	-0.0046	&	-0.0005	&	[2]\\
53100.4600	&	0.0011	&	-14839.0	&	-0.0152	&	-0.0030	&	0.0011	&	[2]\\
53137.4122	&	0.0009	&	-14741.0	&	-0.0178	&	-0.0058	&	-0.0012	&	[2]\\
53137.4163	&	0.0010	&	-14741.0	&	-0.0137	&	-0.0017	&	0.0029	&	[2]\\
53137.4166	&	0.0008	&	-14741.0	&	-0.0134	&	-0.0014	&	0.0032	&	[2]\\
54513.4207	&	0.0006	&	-11092.0	&	-0.0075	&	-0.0007	&	-0.0011	&	[3]\\
54513.6131	&	0.0008	&	-11091.5	&	-0.0037	&	0.0031	&	0.0028	&	[3]\\
55115.4555	&	--	&	-9495.5	&	0.0049	&	0.0099	&	0.0020	&	[4]\\
55276.4695	&	0.0009	&	-9068.5	&	0.0019	&	0.0064	&	-0.0021	&	[5]\\
55277.4074	&	0.0010	&	-9066.0	&	-0.0030	&	0.0016	&	-0.0069	&	[5]\\
55285.5224	&	0.0009	&	-9044.5	&	0.0046	&	0.0092	&	0.0007	&	[5]\\
55289.4804	&	0.0012	&	-9034.0	&	0.0031	&	0.0077	&	-0.0008	&	[5]\\
55765.3665	&	0.0003	&	-7772.0	&	0.0025	&	0.0059	&	0.0002	&	[6]\\
55774.4168	&	0.0002	&	-7748.0	&	0.0027	&	0.0060	&	0.0005	&	[6]\\
55957.4895	&	0.0010	&	-7262.5	&	-0.0014	&	0.0015	&	-0.0015	&	[7]\\
58690.8294	&	0.0001	&	-14.0	&	0.0086	&	0.0086	&	0.0001	&	[2]\\
58690.6406	&	0.0001	&	-14.5	&	0.0083	&	0.0083	&	-0.0002	&	[2]\\
58731.3663	&	0.0001	&	93.5	&	0.0084	&	0.0084	&	0.0000	&	[2]\\
58731.5551	&	0.0001	&	94.0	&	0.0086	&	0.0086	&	0.0002	&	[2]\\
58853.5425	&	0.0001	&	417.5	&	0.0077	&	0.0077	&	0.0000	&	[2]\\
58853.7313	&	0.0001	&	418.0	&	0.0079	&	0.0079	&	0.0002	&	[2]\\
58883.3323	&	0.0001	&	496.5	&	0.0075	&	0.0075	&	-0.0001	&	[2]\\
58883.5211	&	0.0001	&	497.0	&	0.0076	&	0.0076	&	0.0001	&	[2]\\
58908.4085	&	0.0001	&	563.0	&	0.0072	&	0.0072	&	-0.0001	&	[2]\\
58908.5971	&	0.0001	&	563.5	&	0.0072	&	0.0072	&	-0.0001	&	[2]\\
59021.3459	&	0.0001	&	862.5	&	0.0064	&	0.0065	&	0.0002	&	[2]\\
59021.5346	&	0.0001	&	863.0	&	0.0065	&	0.0066	&	0.0003	&	[2]\\
59403.1432	&	0.0001	&	1875.0	&	0.0010	&	0.0012	&	0.0004	&	[2]\\
59403.3318	&	0.0001	&	1875.5	&	0.0011	&	0.0013	&	0.0004	&	[2]\\
59431.0473	&	0.0001	&	1949.0	&	0.0005	&	0.0007	&	0.0003	&	[2]\\
59431.2359	&	0.0001	&	1949.5	&	0.0006	&	0.0008	&	0.0004	&	[2]\\
59580.7490	&	0.0001	&	2346.0	&	-0.0033	&	0.0000	&	0.0002	&	[2]\\
59580.9372	&	0.0001	&	2346.5	&	-0.0036	&	-0.0003	&	-0.0002	&	[2]\\
59606.3901	&	0.0001	&	2414.0	&	-0.0030	&	-0.0027	&	-0.0003	&	[2]\\
59606.5787	&	0.0001	&	2414.5	&	-0.0029	&	-0.0026	&	-0.0002	&	[2]\\
59610.5381	&	0.0001	&	2425.0	&	-0.0030	&	-0.0027	&	-0.0003	&	[2]\\
59610.7267	&	0.0001	&	2425.5	&	-0.0029	&	-0.0026	&	-0.0002	&	[2]\\
59635.4258	&	0.0001	&	2491.0	&	-0.0032	&	-0.0029	&	-0.0001	&	[2]\\
59635.6142	&	0.0001	&	2491.5	&	-0.0033	&	-0.0029	&	-0.0001	&	[2]\\
\hline\hline
\\[.5pt]
\multicolumn{7}{l}{$^1$ The minimum times are in the barycentric dynamical time system (BJD).}
\end{tabular}


\begin{tablenotes}
      \small
      \item Notes: [1] \cite{Pych04}; [2] Present work; [3] \cite{Brat08}; [4] \cite{Kju2010}; [5] \cite{Lia2010}; [6] \cite{Soy2017}; [7] \cite{Hub2013}.
    \end{tablenotes}
\end{threeparttable}
\end{table}


\begin{table} 
\begin{center}
\vspace*{-3.mm}
\footnotesize
\caption{Parameters derived from (\textit{O-C}) analysis.\label{tbl-3}}
\vspace*{-3.mm}
\begin{tabular}{lcc}
      \hline      \hline
       Parameter & Value  & Error \\
      \hline
Parabolic behavior related & & \\
  \hline
      $T_0 $ [BJD] & 2458696.1005 & 0.0001 \\
     $P_{orb}$ [days] & 0.37708906 & 0.00000008 \\
     $dP/dt$ [days/year] & $-1.12 \times 10^{-7}$ & $0.01 \times 10^{-7}$ \\
    \hline
$3^{rd}$ body related & & \\
  \hline
    $a_{12}\,sin\,i$ [AU]  & 1.398 & 0.009 \\
    $e$  & 0.05 & 0.01 \\
     $\omega$ [degree]  & 0.3 & 24.6 \\
    $T_{p}$ [BJD]  & 2462732.2 & 213.7 \\
    $P_{mod}$ [year]  & 9.03 & 0.02 \\
    $A$ [days]  & 0.0081 & 0.0001 \\
     $f(M_3)$  [$ M_{\odot}$]  & 0.03352 & 0.00001 \\
     $M_3(i=90^{\circ})$  [$ M_{\odot}$]  & 0.61145 & 0.00007 \\
     $M_3(i=60^{\circ})$  [$ M_{\odot}$]  & 0.72665 & 0.00009 \\
     $M_3(i=30^{\circ})$  [$ M_{\odot}$]  & 1.48124 & 0.00022 \\
    \hline  
     Magnetic activity cycle related & & \\
  \hline
    $\Delta J$ [$\mathrm{g} \mathrm{cm}^2 \mathrm{s}^{-1}$]  & $7.66 \times 10^{47}$ &  \\
    $\Delta \Omega / \Omega$ & $3.08 \times 10^{-3}$ &  \\
    B[kG]  & 22.9 &  \\
  \hline
  \hline
\end{tabular}
\end{center}
\vspace*{-5.mm}
\end{table}


\begin{landscape}
\small
\begin{longtable}{lcclccccccc}
\caption{Some of the W-type contact binaries with a period decrease.\label{tbl-4}}
 \\
\hline \hline
\\[0.5pt]
       Name	& $P_{orb}$ & $q$ & $M_1$ & $M_2$ &	$dP/dt$ & 	$dM/dt$ &	Cylic & $P_{cyc}$ & LTTE  & Ref. \\
      	& (days) & & M$_{\odot}$ & M$_{\odot}$ & ($\times 10^{-7} \mathrm{d}  \mathrm{yr}^{-1}$)  & ($\times 10^{-7} \mathrm{M}_{\odot} \mathrm{yr}^{-1}$) & & (years) &  &  \\
\\[0.3pt]
\hline
\endfirsthead
\caption{Continued.}\\
\\[0.5pt]
\hline
\\[0.3pt]
       Name	& $P_{orb}$ & $q$ & $M_1$ & $M_2$ &	$dP/dt$ & 	$dM/dt$ &	Cylic & $P_{cyc}$ & LTTE  & Ref. \\
      	& (days) & & M$_{\odot}$ & M$_{\odot}$ & ($\times 10^{-7} \mathrm{d}  \mathrm{yr}^{-1}$)  & ($\times 10^{-7} \mathrm{M}_{\odot} \mathrm{yr}^{-1}$) & & (years) &  &  \\
\\[0.3pt]
\hline
\\[0.3pt]
\endhead
\\[0.3pt]
\hline
\endfoot
\\[0.3pt]
\hline\hline
\\[2pt]
\multicolumn{11}{l}{\hbox to 0pt{\parbox{0.95\linewidth}{\small
      Notes: [1] \cite{Yang2003} ; [2] \cite{Liu2015}; [3] \cite{Stat2021};[4] \cite{Sarot2019aug} ;[5] \cite{Yang2003} ;[6] \cite{Djurasevic2011}; [7] \cite{Yang2011}; [8] \cite{Zhou2020} ; [9] \cite{Xia2018} ; [10] \cite{Kriwat2013} ; [11] \cite{Yang2009} ; [12] \cite{Mitnyan2018} ; [13] \cite{Zhu2019} ;[14] \cite{Lee2009} ;[15] \cite{He2012}  ;[16] \cite{Sarot19apr} ;[17] \cite{Zhang2016} ;[18] \cite{Li2021} ;[19] \cite{Zhu2014} ;[20] \cite{Chris2011} ;[21] \cite{Wang2017} ;[22] \cite{Chris2012} ;[23] \cite{Qian2005} ;[24] \cite{Lu2020} ;[25] \cite{Qian2003};[26] \cite{Qian2007} ;[27] \cite{Liu2011};[28] \cite{Kim2003};[29] \cite{Qian2013};[30] \cite{Yang2008};[31] \cite{Qian2005b} ;[32] \cite{Pych2004};[33] \cite{Maceroni1982};[34] \cite{Liu2006} ;[35] \cite{Zhou2018} ;[36] \cite{Zhu2005} .
}}}
\endlastfoot
CC Com	    &	0.2207 &	0.52 & 0.79 & 0.41 & -0.40  & 0.51 & yes & 16.1 & yes & [1]\\
V1104 Her	&	0.2279 & 	0.63 & 0.74 & 0.46 & -0.29 & 0.52 & yes & 8.28 & yes & [2,3]\\
YZ Phe	    &	0.2347 & 	0.38 & 0.74 & 0.28 & -0.26 & 0.17 & yes & 40.76 & yes & [4]\\
RW Com	    &	0.2370 & 	0.48 & 0.80 & 0.38 & -0.43 & 0.44 & yes & 13.70 & yes & [5,6]\\
EI CVn	    &	0.2608 &	0.46 & 0.63 & 0.29 & -3.11 & 1.03 & unclear & 4.96 & - & [7]\\
V1197 Her	&	0.2627 &	0.38 & 0.77 & 0.30 & -2.58 & 1.61 & unclear & - & - & [8]\\
EH CVn  	&	0.2636 &	0.30 & 0.67 & 0.20 & -0.52 & 0.19 & unclear & - & - & [9]\\
GV Leo  	&	0.2667 &	0.19 & 1.09 & 0.19 & -4.95 & 1.42 & unclear & - & - & [10]\\
BM UMa	    &	0.2712 &	0.54 & 0.92 & 0.50 & -0.75 & 1.01 & yes & 30.8 & unclear & [11]\\
VW Cep	    &	0.2787 &	0.30 & 1.13 & 0.34 & -1.69 & 0.98 & yes & 7.62 & yes & [12]\\
V1005 Her	&	0.2789 &	0.30 & 0.92 & 0.98 & -1.59 & 0.76 & yes & 18.1 & yes & [13]\\
BX Peg	    &	0.2804 &	0.37 & 1.02 & 0.38 & -0.98 & 0.71 &	yes	& 16 & unclear & [14]\\
V524 Mon   &	0.2836 &	0.48 & 0.99 &      & -0.002 & 0.002 &	yes	& 23.9 & unclear & [15]\\
RW Dor     &	0.2854 &	0.63 & 0.97 & 0.61  & -0.14 & 0.27 &	unclear	& 49.9 & - & [16]\\
LO Com     &	0.2864 &	0.40 & 0.79 & 0.32  & -1.18 & 0.74 &	unclear	& - & - & [17]\\
V842 Cep     &	0.2889 &	0.44 & 0.76 & 0.33  & -1.50 & 1.01 &	unclear	& - & - & [18]\\
EP Cep     &	0.2897 &	0.15 & 0.73 & 0.11  & -3.73 & 0.56 &	unclear	& - & - & [19]\\
TZ Boo     &	0.2974 &	0.21 & 0.72 & 0.11  & -0.21 & 0.03 &	yes	& 31.2 & unclear & [20]\\
V2284 Cyg   &	0.3069 &	0.35 & 0.86 & 0.30  & -2.67	& 1.34 &	yes & 2.06 & yes & [21]\\
TY Boo	    &	0.3171 &	0.47 & 1.21 & 0.57 & -3.6	& 0.41 &	yes	& 58.9 & yes & [22]\\
FG Hya	    &	0.3278 &	0.11 & 1.44 & 0.16 & -1.96	& 0.36 &	yes	& 36.4 & unclear & [23]\\
V781 Tau    &	0.3449 &	0.45 & 1.06 & 0.43 & -0.32	& 0.22 &	yes	& 30.8 & unclear & [24]\\
V417 Aql    &	0.3700 &	0.37 & 1.40 & 0.50 & -0.55	& 0.39 &	yes	& 42.4 & unclear & [25]\\
AP Leo     &	0.3703 &	0.30 & 1.47 & 0.44 & -1.08	& 0.61 &	yes	& 22.4 & unclear & [26]\\
V396 Mon   &	0.3963 &	0.39 & 0.92 & 0.36 & -0.86	& 0.51 &	yes	& 42.4 & yes & [27]\\
SS Ari     &	0.4060 &	0.36 & 1.30 & 0.40 & -1.56	& 0.74 &	yes	& 39.7 & unclear & [28]\\
MR Com	   &	0.4127 &	0.26 & 1.40 & 0.36 & -5.30	& 2.07 &	yes	& 10.1 & yes & [29]\\
BS Cas	   &	0.4408 &	0.28 & 1.20 &      & -1.51	& 0.54 &	yes	& 13.2 & unclear & [30]\\
TV Mus	   &	0.4457 &	0.17 & 1.35 & 0.22 & -2.16	& 0.42 &	yes	& 29.1 & unclear & [31]\\
V502 Oph    &	0.4534 &	0.34 & 1.42 & 0.48 & -1.68	& 0.88 &	yes	& 23 & yes & [32,33,34]\\
GU Ori      &	0.4706 &	0.43 & 1.05 & 0.45 & -0.62	& 0.35 &	unclear	& -  & - & [35]\\
IK Per      &	0.6760 &	0.19 & 1.99 & 0.34 & -2.52	& 0.51 &	yes	& 50.5  & yes & [36]\\
\end{longtable}
\end{landscape}

\begin{figure}
	\includegraphics[width=1.0\columnwidth]{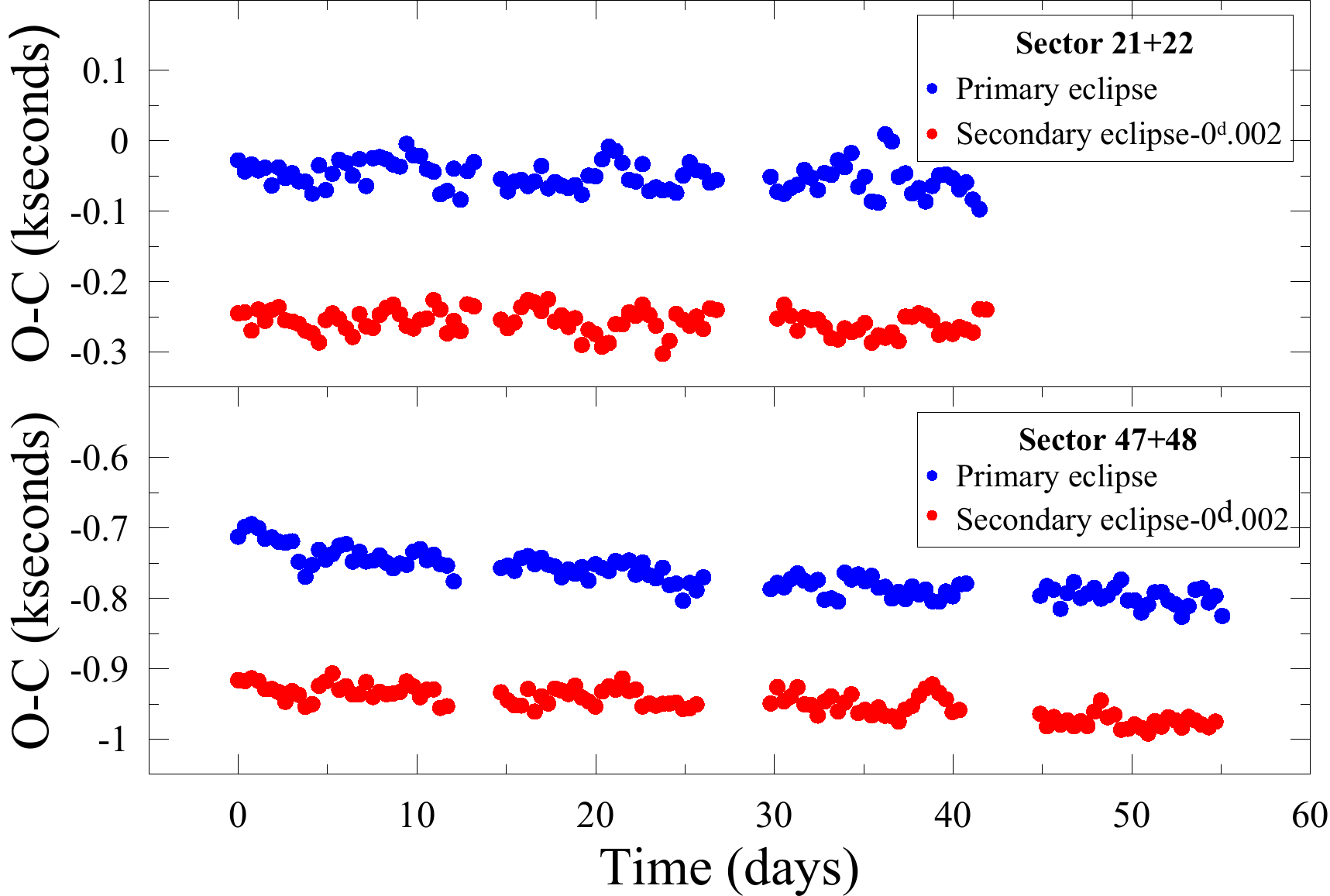}
    \caption{Time-dependent variations of the \textit{O-C} curves for the primary and secondary eclipses (vertically displaced for clarity) are shown as blue and red colors for Sector 21, 22 (top panel), and Sector 47, 48 (bottom panel), respectively.}
    \label{Fig.4 }
\end{figure}

\begin{figure}
	\includegraphics[width=1.0\columnwidth]{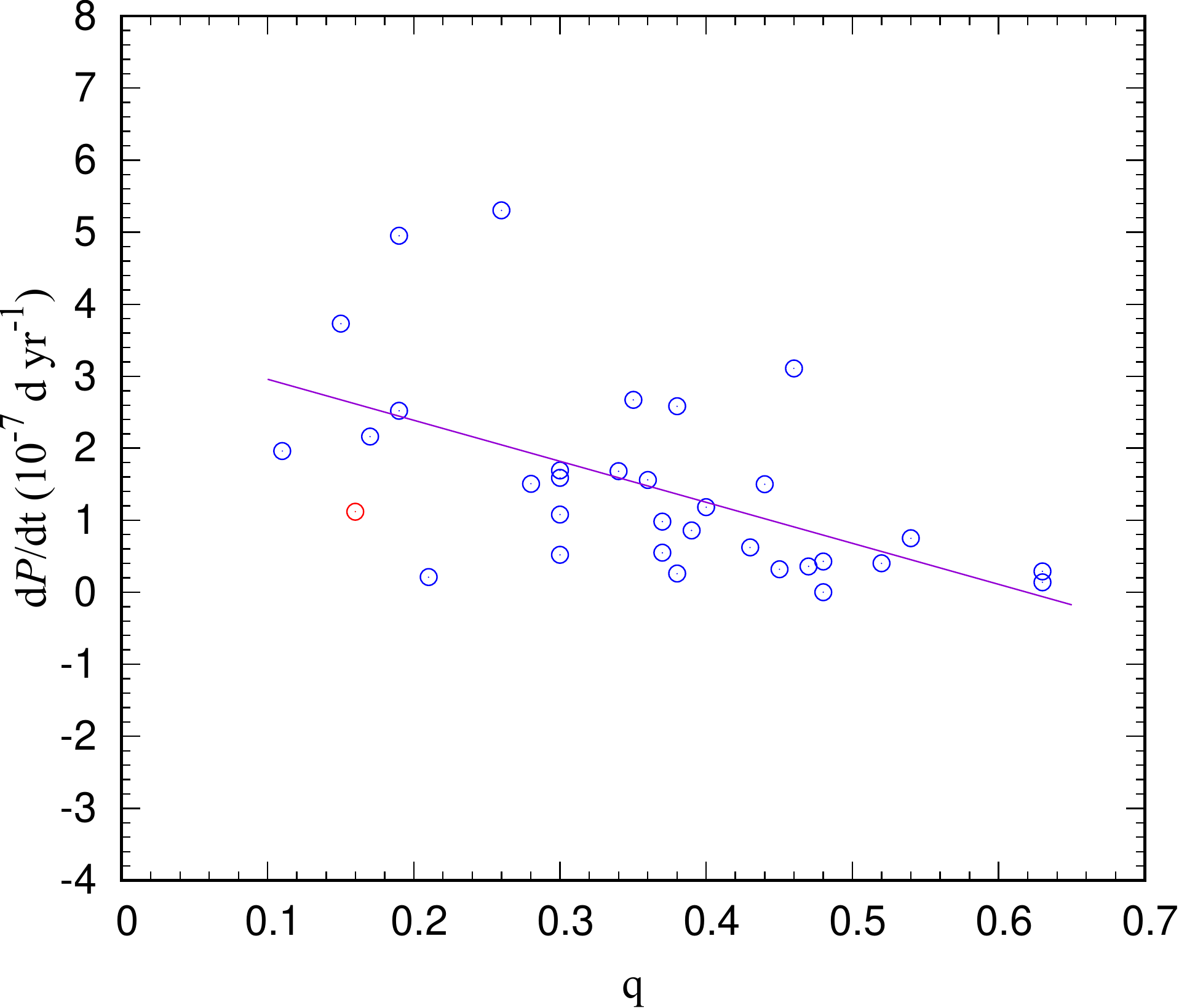}
    \caption{ A plot of the period change range versus the mass ratio ($dP/dt$- q) for some W-type systems (blue circles) listed in Tab.\,\ref{tbl-4}. For comparison, the parameters of TU UMi (red circles) are also plotted. The fitted curve is given in Eq. 4.}
    \label{Fig.5 }
\end{figure}

\bibliography{DZC}

\end{document}